\documentclass[conference]{IEEEtran}
\pdfoutput=1
\usepackage{fancyhdr}
\fancypagestyle{IEEEtitlepagestyle}{
  \fancyhf{}
  \fancyfoot[C]{\sffamily\footnotesize\thepage}
  \fancyfoot[L]{\sffamily\scriptsize Copyright held by the owner/author(s). Presented at\\
3rd International Workshop on Overlay Architectures for FPGAs (OLAF2017)
\\Monterey, CA, USA, Feb. 22, 2017.}

}
\makeatletter
\newif\if@restonecol
\makeatother

\usepackage[ruled]{algorithm2e}
\usepackage{multirow}
\usepackage[center]{caption}
\usepackage{setspace}
\usepackage{amsmath}
\usepackage{multirow}
\usepackage{todonotes}
\usepackage{tikz}
\usepackage{xcolor}
\usepackage{pgfplots, pgfplotstable}
\usepackage{amsmath}
\usepackage[ruled]{algorithm2e}
\usepackage{multirow}
\usepackage{booktabs}
\usepackage{siunitx}
\usepackage{calc}
\usepackage{array}
\usepackage{caption}
\usepackage{rotating}
\usepackage{url}
\usepackage{pgf,tikz}
\usetikzlibrary{trees}
\usetikzlibrary{snakes,arrows,shapes}

\usepackage{amsmath}
\usepackage{mdwlist}
\usepackage[normal]{subfigure}

\usepackage{graphicx}
\usepackage{epstopdf}
\usepackage{epsfig}
\usepackage{listings}
\usepackage{xcolor}
\usepackage{dot2texi}
\usepackage{graphicx,stfloats}

\usepackage{graphicx,epstopdf}
\pgfplotsset{compat=newest}

\begin{document}
\title{Resource-Aware Just-in-Time OpenCL Compiler for Coarse-Grained FPGA Overlays}

\author{\IEEEauthorblockN{Abhishek~Kumar~Jain, Douglas~L.~Maskell}
	\IEEEauthorblockA{School of Computer Science and Engineering\\
		Nanyang Technological University, Singapore \\ Email: \{abhishek013, asdouglas\}@ntu.edu.sg}
	\and
	\IEEEauthorblockN{Suhaib~A.~Fahmy}
	\IEEEauthorblockA{School of Engineering\\
		University of Warwick, Coventry, UK \\ Email: s.fahmy@warwick.ac.uk}}

\maketitle
\setlength{\baselineskip}{9.7pt}

\begin{abstract}
FPGA vendors have recently started focusing on OpenCL for FPGAs because of its ability to leverage the parallelism inherent to heterogeneous computing platforms. OpenCL allows programs running on a host computer to launch accelerator kernels which can be compiled at run-time for a specific architecture, thus enabling portability. However, the prohibitive compilation times (specifically the FPGA place and route times) are a major stumbling block when using OpenCL tools from FPGA vendors. The long compilation times mean that the tools cannot effectively use just-in-time (JIT) compilation or runtime performance scaling. Coarse-grained overlays represent a possible solution by virtue of their coarse granularity and fast compilation. In this paper, we present a methodology for run-time compilation of OpenCL kernels to a DSP block based coarse-grained overlay, rather than directly to the fine-grained FPGA fabric. The proposed methodology allows JIT compilation and on-demand resource-aware kernel replication to better utilize available overlay resources, raising the abstraction level while reducing compile times significantly. We further demonstrate that this approach can even be used for run-time compilation of OpenCL kernels on the ARM processor of the embedded heterogeneous Zynq device.
\end{abstract}

\IEEEpeerreviewmaketitle

%-------------------------------------------------------------------------
% Section 1
%-------------------------------------------------------------------------
\section{Introduction}
\label{sec1_introduction}
The open computing language (OpenCL) has become an industry standard for parallel programming on a range of heterogeneous platforms including CPUs, GPUs, FPGAs, and other accelerators. OpenCL also has the benefit of being portable across architectures without changes to algorithm source code~\cite{chen2012invited}. In OpenCL, parallelism is explicitly specified by the programmer, and compilers can use system information at runtime to scale the performance of the application by executing multiple replicated copies of the application kernel in hardware~\cite{gao2014characterization}.
That is, while OpenCL supports both online and offline compilation, application kernels in OpenCL are intended to be compiled at run-time~\cite{stone2010opencl}, so that applications are more portable across various platforms. This online compilation of kernels is referred to as just-in-time (JIT) compilation.

FPGA vendors have recently released OpenCL based tools (Altera OpenCL and Xilinx SDAccel) to bridge the gap between the expressiveness of sequential programming languages and the parallel capabilities of the FPGA hardware~\cite{singh2013harnessing}. One reason for this is the introduction of more capable heterogeneous system on chip (SoC) platforms/hybrid FPGAs which tightly couple general purpose processors with high performance FPGA fabrics~\cite{ahmad201616} and provide a more energy efficient alternative to high performance CPUs and/or GPUs within the tight power budget required by high performance embedded systems.
Hence techniques for mapping OpenCL kernels to FPGA hardware have attracted both academic and industrial attention in the last few years~\cite{owaida_synthesis_2011,czajkowski_opencl_2012,shagrithaya_enabling_2013}. 

The design process on FPGAs is somewhat different to that of CPU or GPU based accelerators. Providing OpenCL/high level synthesis (HLS) support for FPGA platforms has helped simplify accelerator design by raising the programming abstraction above RTL. This has allowed accelerator functionality to be described at a higher level to reduce developer effort, enable design portability and enable rapid design space exploration, thus improving productivity, verifiability, and flexibility. 
However, this is just one part of the FPGA design process, and unfortunately, the compilation time for FPGAs is extremely long (hours rather than seconds), limiting the FPGA to fixed off-line accelerator implementations, similar to using pre-compiled kernel binaries on GPUs. This can be a significant liability for software developers who are accustomed to rapid compile times, with a fast turnaround allowing more efficient testing and tuning of accelerator kernels.
It also prevents FPGAs from taking advantage of JIT compilation and dynamic performance scaling using kernel replication~\cite{gao2014characterization} when additional resource becomes available. 
Thus, on FPGA, design iterations are extremely slow, and as the FPGA devices grow in size, and designs occupy larger areas, this problem will continue to worsen, making JIT compilation on FPGA unlikely.

Coarse-grained overlay architectures built on top of the FPGA have emerged as a possible solution to this challenge~\cite{stitt_intermediate_2011,capalija_high-performance_2013,benson2012design,jainadapting2016}, thereby offering a simpler target architecture for the compilation flow, resulting in fast compilation, but at the cost of sometimes significant area and performance overheads.
Recent research in this area has demonstrated ways in which more efficient overlays can be built~\cite{fccm2015-jain,date2016-jain,jain2016deco}. These overlays, coupled with high-level design methods, could address both the design and compilation aspects of the design productivity gap.

In this paper, we first present a methodology for JIT (run-time) compilation of OpenCL kernels targeting an efficient coarse-grained overlay, rather than directly to the FPGA fabric. 
We use a custom automated mapping flow based around the LLVM front-end to provide a rapid, vendor independent, mapping to the overlay.
The methodology benefits from the high level of abstraction afforded by using the OpenCL programming model, while the mapping to the overlay significantly reduces the compilation and load times.
Next, we present a comparison of place and route (PAR) times using vendor tools targeting traditional fine-grained architecture and the proposed overlay approach for a set of benchmarks.
Finally, we demonstrate run-time performance scaling using the concept of on-demand resource-aware OpenCL kernel replication running entirely on the embedded processor in the Xilinx Zynq.

%-------------------------------------------------------------------------
% Section 2
%-------------------------------------------------------------------------
\section{Motivation and Related Work}
\label{sec2_motivation}
Most vendor OpenCL tools simply generate intermediate RTL that must still be mapped through the FPGA backend flow. 
Unfortunately, the size and fine granularity of modern FPGA fabrics means that PAR times are very long, thereby preventing runtime compilation of OpenCL kernels.
PAR times for coarse-grained architectures, such as coarse-grained reconfigurable arrays (CGRAs), are significantly reduced. 
However, CGRAs implemented as ASIC devices~\cite{ebeling_rapid_1996,liang_smartcell:_2009,mei_adres:_2003} have not been successful because functional units (FUs) are often too application specific to be efficient and useful for a wide enough range of applications~\cite{liang_smartcell:_2009}.
In contrast, coarse-grained reconfigurable architectures implemented on top of commercial FPGAs can be easily tuned to a particular application, allowing the FU and interconnect to be adapted according to application requirements~\cite{jain2016coarse}.
Unfortunately, many of the overlay designs proposed in the literature do not consider the FPGA architecture to a significant extent, and as a result, suffer from significant area and performance overheads.
An efficient overlay architecture, built around the capabilities of the architecture, coupled with a high level design approach, such as OpenCL, would tackle the two key issues in design productivity.

OpenCL targeting the TILT overlay~\cite{rashid2014comparing} was suggested as an alternative to the Altera OpenCL tool (which generates a heavily pipelined, spatial design which maximizes throughput at the cost of significant resource usage) when a lower throughput is adequate.
TILT uses a weaker form of application customization by varying the mix of pre-configured standard FUs and optionally generating application-dependent custom units.
However, as each application requires that the TILT overlay be recompiled, a hardware context switch (referred to as a kernel update in the paper) takes on average 38 seconds.
Additionally, the Altera OpenCL HLS implementation of the benchmark application was 2$\times$ more efficient in terms of throughput per unit area.

In~\cite{coole2014fast}, the authors used overlays having one dedicated functional unit for each OpenCL kernel operation.
Five different relatively small sized overlays (2 floating point and 3 fixed point) were designed, each specialized for a specific set of kernels, and implemented on a Xilinx Virtex-6 FPGA (XC6VCX130T) running at frequencies ranging from 196\,MHz to 256\,MHz.
When the overlay residing on the FPGA fabric did not support a kernel, it was proposed to reconfigure the FPGA fabric at runtime with a different overlay which supported the new kernel.
This was needed because different applications require different sized overlays, with an overlay large enough to satisfy the resource requirements of the largest kernel being heavily underutilized when a small kernel is mapped to the overlay.

In our work, instead of compiling OpenCL kernels onto relatively small kernel set-specific overlays, we compile replicated instances of kernels onto a large overlay to achieve effective utilization of resources. The size of the overlay on the fabric depends on the available resources after any other logic is mapped.
In the case where there is minimal, or no, requirement for other logic we can completely fill the fabric with the largest possible overlay. Hence, the overlays we consider can have different sizes and FU types, with this information being exposed by the OpenCL runtime to the compiler, which can then perform on-demand resource-aware kernel replication to effectively utilize the available overlay resources.

Additionally, in the context of software/hardware systems on hybrid FPGAs, the reconfiguration time required to load a hardware accelerator onto the FPGA, and not just the compile time, is also a significant factor in exploiting the acceleration benefits and the virtualization feature of the FPGA hardware~\cite{cloudcom2015-fahmy,jsps2014-jain}.
Hence, the fast configuration afforded by overlays is important in such systems where software and hardware execution is tightly coupled.
Furthermore, overlays offer the benefit of portability, whereby the same application can be loaded onto identical overlays on different physical devices without re-compilation. That is, overlays offer the potential to support FPGA usage models where programmability, abstraction, resource sharing, fast compilation, and design productivity are critical issues.

%-------------------------------------------------------------------------
% Section 3
%-------------------------------------------------------------------------
\section{Compiling Kernels to the Overlay}
\label{sec4_compilation}
The overlay we use in this paper to demonstrate the compilation flow is a spatially configured array of functional units interconnected using a programmable interconnect architecture, as described in~\cite{fccm2015-jain,date2016-jain}.
The programmable FU executes arithmetic operations and data is transferred over a dedicated, but programmable, point-to-point link between the FUs.
That is, both the FU and the interconnect are unchanged while a compute kernel executes.
This results in a fully pipelined, throughput oriented programmable datapath executing one kernel iteration per clock cycle, thus having an initiation interval (II) between kernel data packets of one.
This is different from many of the CGRAs in the literature which are time-multiplexed~\cite{liu2015quickdough,mei_adres:_2003}, where individual FUs have their own set of instructions and memory, and have an II greater than one.

\begin{figure}[!h]
	\centering
	\subfigure[Block diagram.]{
		\centering
		\includegraphics[width=2.8cm]{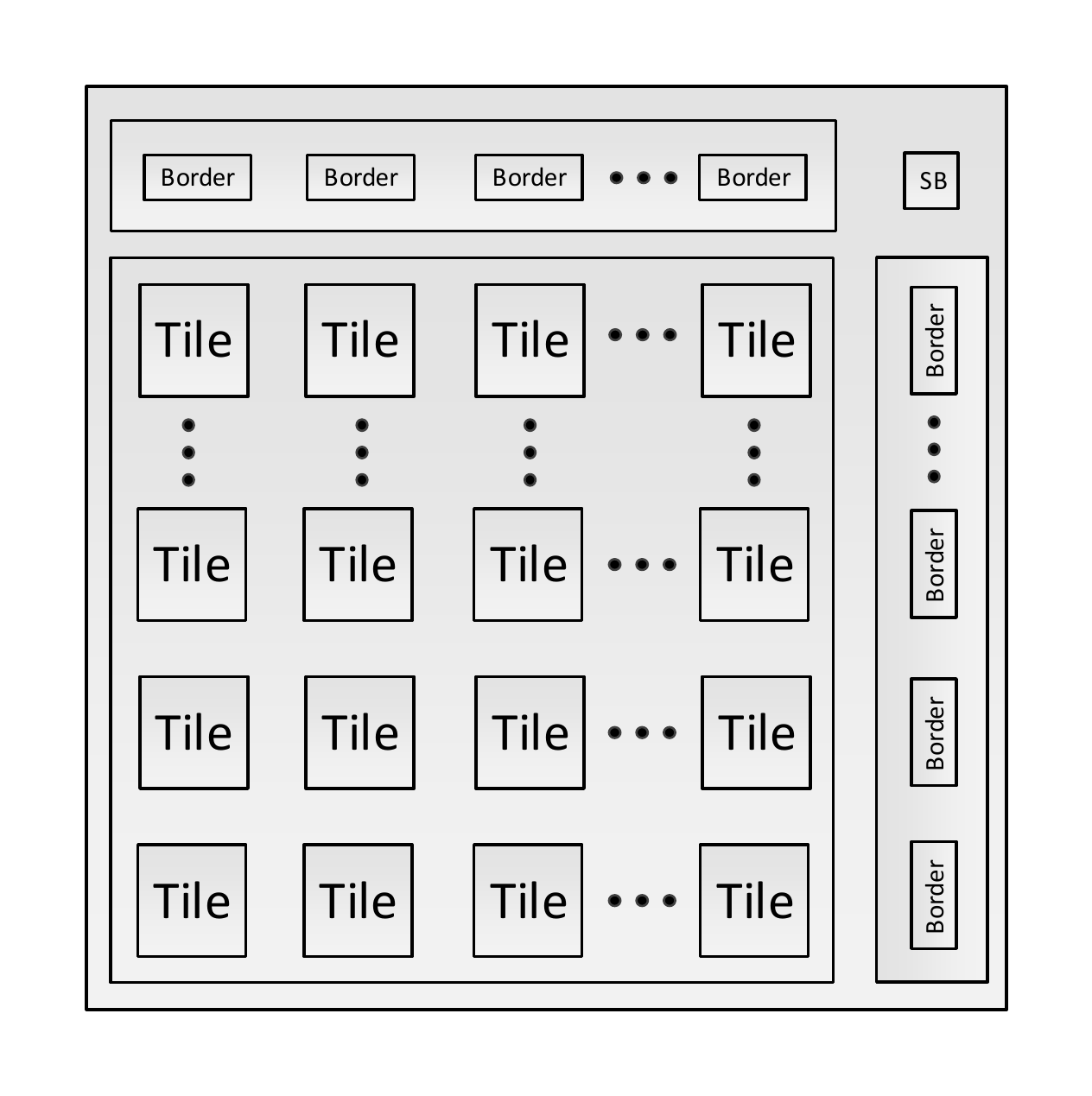}
		\label{overlay_fccm}
	}
	\subfigure[Tile architecture.]{
		\centering
		\includegraphics[width=3.5cm]{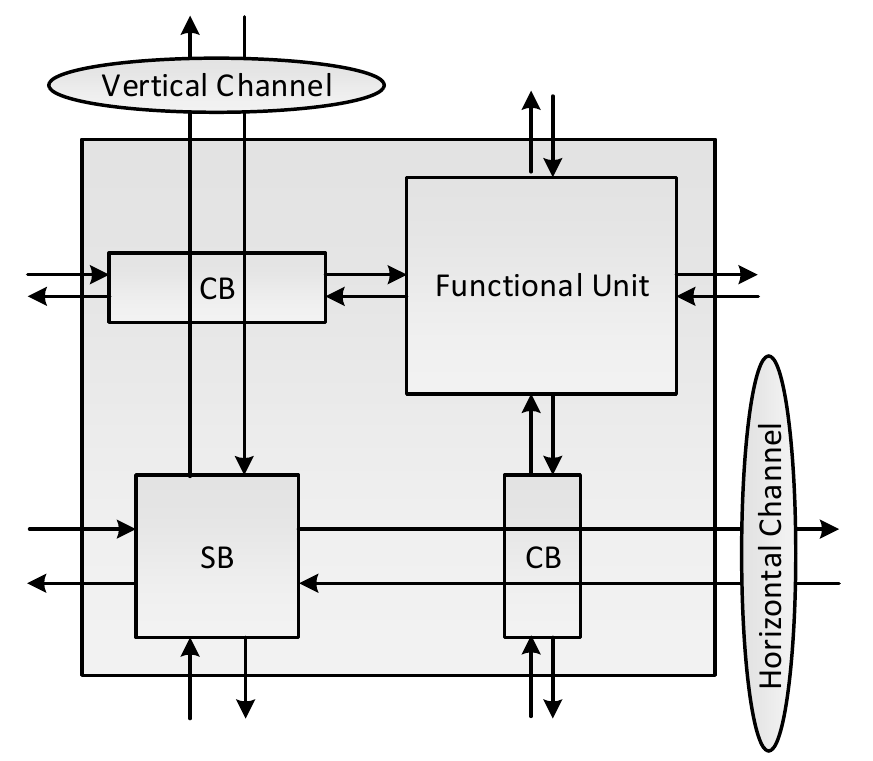}
		\label{tile_fccm}
	}
	\caption[Overlay architecture.]{Overlay architecture~\cite{fccm2015-jain,date2016-jain}.}
	\label{arch}
\end{figure}

The programmability of spatially-configured overlays comes at the cost of area and performance overheads, which are magnified if the overlay is designed without consideration for the underlying FPGA architecture.
The overlay we use in this paper consists of a traditional island-style topology, arranged as a virtual homogeneous two-dimensional array of tiles, where each tile consists of routing resources (one switch box and 2 connection boxes) and a DSP Block based FU, as shown in Fig.~\ref{arch}. 
Using DSP blocks in the FU with deep pipelining results in a high throughput and improved area efficiency~\cite{fccm2015-jain,date2016-jain}.
Configurable shift registers are placed at each DSP input for balancing pipeline latencies. 
When mapped to a Xilinx Zynq XC7Z020 device, an overlay with two DSPs per FU can provide a peak throughput of 115 GOPS~\cite{date2016-jain} (throughputs of up to 912 GOPS have been reported when mapped to a more capable Virtex 7 device~\cite{date2016-jain}).
Additional details on the overlay architecture are available in~\cite{fccm2015-jain,date2016-jain}.

\vspace{-3mm}

\vspace{-3mm}

The overlay design and implementation still requires the conventional hardware design flow using vendor tools.
However, this process is done offline only once, and so does not impact the kernel implementation of an application.
This is different from some of the other overlays which require the overlay itself to be adapted to the kernel being mapped, and hence lose the benefit of fast compilation~\cite{liu2015quickdough}.

\begin{figure}[!h]
	\centering
	\includegraphics[width=7.0cm]{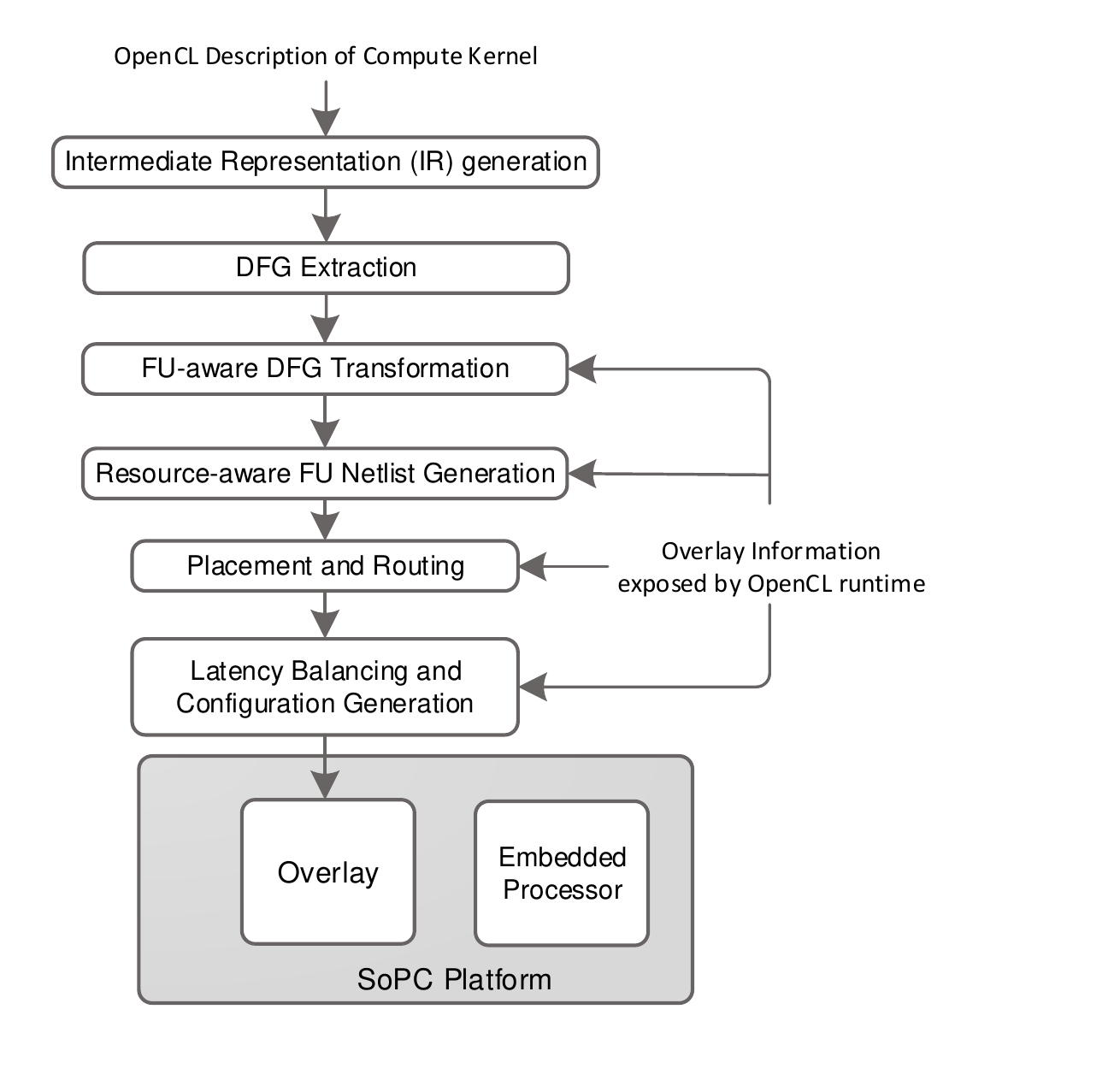}
	\caption{Automated mapping flow.}
	\label{flow}
\end{figure}

Instead of compiling high level application kernels to RTL and then generating a bitstream using the vendor tools, we have developed our own automated mapping flow to provide a rapid, vendor independent, mapping to the overlay.
The mapping process comprises LLVM Intermediate Representation (IR) generation for an OpenCL kernel using Clang, IR optimization using LLVM optimization passes, DFG extraction from the IR, mapping of the DFG nodes onto the overlay FUs, VPR compatible FU netlist generation, placement and routing of the FU netlist onto the overlay, latency balancing and finally, configuration generation.
The automated overlay mapping flow is shown in Fig.~\ref{flow} and is described in detail below by demonstrating the step by step process of compiling a simple example OpenCL kernel. 

\vspace{-3mm}

\vspace{-4mm}

\lstset { %
	language=C,
	backgroundcolor=\color{white}, 
	basicstyle=\ttfamily\tiny,
	keywordstyle=\color{blue}\ttfamily,
	stringstyle=\color{red}\ttfamily,
	commentstyle=\color{green}\ttfamily,
	breaklines=true	
}
\lstset{framesep=-5pt, xleftmargin=-5pt}
\begin{table}[!h]
	\caption{Compute Kernel Code Descriptions}
	\centering
	\label{code.tbl}
	\begin{tabular}{l}
		\toprule
		\multicolumn{1}{c}{(a) OpenCL Description of the Kernel}\\ % Assembly with Loopback Optimization
		\midrule
		%    \hspace{-0.2in}
\begin{lstlisting}[columns=fullflexible, language=C]
__kernel void example_kernel(__global int *A, __global int *B)
{
int idx = get_global_id(0);
int x = A[idx];
B[idx] = (x*(x*(16*x*x-20)*x+5)) ;
}
\end{lstlisting}		
\\
\midrule		
\multicolumn{1}{c}{(b) Intermediate Representation (IR) of the Kernel}
\\				
\midrule
\begin{lstlisting}
%0:
%1 = alloca i32*, align 4
%2 = alloca i32*, align 4
%idx = alloca i32, align 4
%x = alloca i32, align 4
store i32* %A, i32** %1, align 4
store i32* %B, i32** %2, align 4
%3 = call i32 bitcast (i32 (...)* @get_global_id to i32 (i32)*)(i32 0)
store i32 %3, i32* %idx, align 4
%4 = load i32* %idx, align 4
%5 = load i32** %1, align 4
%6 = getelementptr inbounds i32* %5, i32 %4
%7 = load i32* %6
store i32 %7, i32* %x, align 4
%8 = load i32* %x, align 4
%9 = load i32* %x, align 4
%10 = load i32* %x, align 4
%11 = mul nsw i32 16, %10
%12 = load i32* %x, align 4
%13 = mul nsw i32 %11, %12
%14 = sub nsw i32 %13, 20
%15 = mul nsw i32 %9, %14
%16 = load i32* %x, align 4
%17 = mul nsw i32 %15, %16
%18 = add nsw i32 %17, 5
%19 = mul nsw i32 %8, %18
%20 = load i32* %idx, align 4
%21 = load i32** %2, align 4
%22 = getelementptr inbounds i32* %21, i32 %20
store i32 %19, i32* %22
ret void
\end{lstlisting}
\\
\midrule		
\multicolumn{1}{c}{(c) Optimized IR of the Kernel}
\\				
\midrule
\begin{lstlisting}
%0:
%1 = call i32 bitcast (i32 (...)* @get_global_id to i32 (i32)*)(i32 0)
%2 = getelementptr inbounds i32* %A, i32 %1
%3 = load i32* %2
%4 = mul nsw i32 16, %3
%5 = mul nsw i32 %4, %3
%6 = sub nsw i32 %5, 20
%7 = mul nsw i32 %3, %6
%8 = mul nsw i32 %7, %3
%9 = add nsw i32 %8, 5
%10 = mul nsw i32 %3, %9
%11 = getelementptr inbounds i32* %B, i32 %1
store i32 %10, i32* %11
ret void
\end{lstlisting}
\\				
\bottomrule
\end{tabular}
\end{table}

\lstset { %
	language=C,
	backgroundcolor=\color{white},
	basicstyle=\ttfamily\tiny,
	keywordstyle=\color{blue}\ttfamily,
	stringstyle=\color{red}\ttfamily,
	commentstyle=\color{green}\ttfamily,
	breaklines=true	
}
\lstset{framesep=-5pt, xleftmargin=-5pt}
\begin{table}[!h]
\caption{Compute Kernel DFG Descriptions}
\centering
\label{dfg.tbl}
\begin{tabular}{l}
\toprule
\multicolumn{1}{c}{(a) DFG Description of the Kernel}
\\				
\midrule
\begin{lstlisting}
digraph example_kernel {
N8 [ntype="operation", label="add_Imm_5_N8"];
N9 [ntype="outvar", label="O0_N9"];
N1 [ntype="invar", label="I0_N1"];
N2 [ntype="operation", label="mul_N2"];
N3 [ntype="operation", label="mul_N3"];
N4 [ntype="operation", label="mul_Imm_16_N4"];
N5 [ntype="operation", label="mul_N5"];
N6 [ntype="operation", label="mul_N6"];
N7 [ntype="operation", label="sub_Imm_20_N7"];
N8 -> N2;
N1 -> N5;
N1 -> N6;
N1 -> N2;
N1 -> N3;
N1 -> N4;
N2 -> N9;
N3 -> N6;
N4 -> N5;
N5 -> N7;
N6 -> N8;
N7 -> N3;
}
\end{lstlisting}
\\		
\midrule
\multicolumn{1}{c}{(b) FU-aware DFG Description of the Kernel}
\\				
\midrule
\begin{lstlisting}
digraph example_kernel {
N7 [ntype="outvar", label="O0_N7"];
N1 [ntype="invar", label="I0_N1"];
N2 [ntype="operation", label="mul_N2"];
N3 [ntype="operation", label="mul_N3"];
N4 [ntype="operation", label="mul_Imm_16_N4"];
N5 [ntype="operation", label="mul_sub_Imm_20_N5"];
N6 [ntype="operation", label="mul_add_Imm_5_N6"];
N1 -> N5;
N1 -> N6;
N1 -> N2;
N1 -> N3;
N1 -> N4;
N2 -> N7;
N3 -> N6;
N4 -> N5;
N5 -> N3;
N6 -> N2;
}
\end{lstlisting}
\\				
\bottomrule
\end{tabular}
\end{table}

\subsection{DFG extraction from a kernel description}
To support OpenCL, we use LLVM, which can extract a DFG from an OpenCL description using the following steps:

\subsubsection{IR generation from an OpenCL kernel}
Given the simple OpenCL kernel shown in Table~\ref{code.tbl}(a), the LLVM compiler front-end (Clang), along with the LLVM disassembler, generates IR, as in Table~\ref{code.tbl}(b).
LLVM optimization passes are then used to generate an optimized LLVM IR, as in Table~\ref{code.tbl}(c).

\pgfdeclarelayer{background}
\pgfdeclarelayer{foreground}
\pgfsetlayers{background,main,foreground}

\begin{figure*}[!t]
	\centering
	\subfigure[DFG extracted from Kernel]{
		\centering
		\begin{tikzpicture}[>=latex',line join=bevel, scale = 0.28]
		\centering
		\tiny
		\node (N8) at (92.0bp,124.5bp) [draw=black,ellipse] {add\_Imm\_5\_N8};
		\node (N9) at (46.0bp,10.5bp) [draw=black,ellipse] {O0\_N9};
		\node (N1) at (112.0bp,466.5bp) [draw=black,ellipse] {I0\_N1};
		\node (N2) at (46.0bp,67.5bp) [draw=black,ellipse] {mul\_N2};
		\node (N3) at (102.0bp,238.5bp) [draw=black,ellipse] {mul\_N3};
		\node (N4) at (112.0bp,409.5bp) [draw=black,ellipse] {mul\_Imm\_16\_N4};
		\node (N5) at (112.0bp,352.5bp) [draw=black,ellipse] {mul\_N5};
		\node (N6) at (102.0bp,181.5bp) [draw=black,ellipse] {mul\_N6};
		\node (N7) at (112.0bp,295.5bp) [draw=black,ellipse] {sub\_Imm\_20\_N7};
		\draw [->] (N1) ..controls (146.27bp,446.62bp) and (178.55bp,423.54bp)  .. (167.0bp,399.0bp) .. controls (160.57bp,385.33bp) and (147.79bp,374.36bp)  .. (N5);
		\draw [->] (N1) ..controls (159.02bp,444.04bp) and (224.0bp,407.12bp)  .. (224.0bp,353.5bp) .. controls (224.0bp,353.5bp) and (224.0bp,353.5bp)  .. (224.0bp,294.5bp) .. controls (224.0bp,243.78bp) and (166.06bp,209.78bp)  .. (N6);
		\draw [->] (N3) ..controls (102.0bp,220.91bp) and (102.0bp,211.14bp)  .. (N6);
		\draw [->] (N4) ..controls (112.0bp,391.91bp) and (112.0bp,382.14bp)  .. (N5);
		\draw [->] (N5) ..controls (112.0bp,334.91bp) and (112.0bp,325.14bp)  .. (N7);
		\draw [->] (N1) ..controls (64.984bp,444.04bp) and (0.0bp,407.12bp)  .. (0.0bp,353.5bp) .. controls (0.0bp,353.5bp) and (0.0bp,353.5bp)  .. (0.0bp,180.5bp) .. controls (0.0bp,144.98bp) and (20.045bp,107.55bp)  .. (N2);
		\draw [->] (N7) ..controls (108.98bp,277.91bp) and (107.21bp,268.14bp)  .. (N3);
		\draw [->] (N1) ..controls (84.093bp,450.73bp) and (65.284bp,437.6bp)  .. (57.0bp,420.0bp) .. controls (31.453bp,365.71bp) and (34.192bp,340.5bp)  .. (57.0bp,285.0bp) .. controls (62.03bp,272.76bp) and (72.121bp,262.06bp)  .. (N3);
		\draw [->] (N8) ..controls (77.444bp,106.1bp) and (68.035bp,94.846bp)  .. (N2);
		\draw [->] (N2) ..controls (46.0bp,49.908bp) and (46.0bp,40.144bp)  .. (N9);
		\draw [->] (N1) ..controls (112.0bp,448.91bp) and (112.0bp,439.14bp)  .. (N4);
		\draw [->] (N6) ..controls (98.983bp,163.91bp) and (97.208bp,154.14bp)  .. (N8);

		\end{tikzpicture}
		\label{dfg}
	}
	\hfill
	\subfigure[FU-aware DFG where FU consists of one DSP block]{
		\centering
		\begin{tikzpicture}[>=latex',line join=bevel, scale = 0.28]
		\tiny%
		\node (N9) at (46.0bp,10.5bp) [draw=black,ellipse] {O0\_N7};
		\node (N1) at (112.0bp,466.5bp) [draw=black,ellipse] {I0\_N1};
		\node (N2) at (46.0bp,67.5bp) [draw=black,ellipse] {mul\_N2};
		\node (N3) at (102.0bp,238.5bp) [draw=black,ellipse] {mul\_N3};
		\node (N4) at (112.0bp,409.5bp) [draw=black,ellipse] {mul\_Imm\_16\_N4};
		\node (N5) at (112.0bp,352.5bp) [draw=black,ellipse] {mul\_sub\_Imm\_20\_N5};
		\node (N6) at (102.0bp,181.5bp) [draw=black,ellipse] {mul\_add\_Imm\_5\_N6};
		
		\draw [->] (N1) ..controls (146.27bp,446.62bp) and (178.55bp,423.54bp)  .. (167.0bp,399.0bp) .. controls (160.57bp,385.33bp) and (147.79bp,374.36bp)  .. (N5);
		\draw [->] (N1) ..controls (159.02bp,444.04bp) and (224.0bp,407.12bp)  .. (224.0bp,353.5bp) .. controls (224.0bp,353.5bp) and (224.0bp,353.5bp)  .. (224.0bp,294.5bp) .. controls (224.0bp,243.78bp) and (166.06bp,209.78bp)  .. (N6);
		\draw [->] (N3) ..controls (102.0bp,220.91bp) and (102.0bp,211.14bp)  .. (N6);
		\draw [->] (N4) ..controls (112.0bp,391.91bp) and (112.0bp,382.14bp)  .. (N5);
		\draw [->] (N5) ..controls (112.0bp,334.91bp) and (112.0bp,325.14bp)  .. (N3);
		\draw [->] (N1) ..controls (64.984bp,444.04bp) and (0.0bp,407.12bp)  .. (0.0bp,353.5bp) .. controls (0.0bp,353.5bp) and (0.0bp,353.5bp)  .. (0.0bp,180.5bp) .. controls (0.0bp,144.98bp) and (20.045bp,107.55bp)  .. (N2);
		\draw [->] (N1) ..controls (84.093bp,450.73bp) and (65.284bp,437.6bp)  .. (57.0bp,420.0bp) .. controls (31.453bp,365.71bp) and (34.192bp,340.5bp)  .. (57.0bp,285.0bp) .. controls (62.03bp,272.76bp) and (72.121bp,262.06bp)  .. (N3);
		\draw [->] (N2) ..controls (46.0bp,49.908bp) and (46.0bp,40.144bp)  .. (N9);
		\draw [->] (N1) ..controls (112.0bp,448.91bp) and (112.0bp,439.14bp)  .. (N4);
		\draw [->] (N6) ..controls (98.983bp,163.91bp) and (97.208bp,154.14bp)  .. (N2);

		\end{tikzpicture}
		\label{fu_dfg_1d}
	}
	\hfill
	\subfigure[FU-aware DFG placed and routed on 5x5 overlay]{
		\centering
		\includegraphics[width=3.5cm]{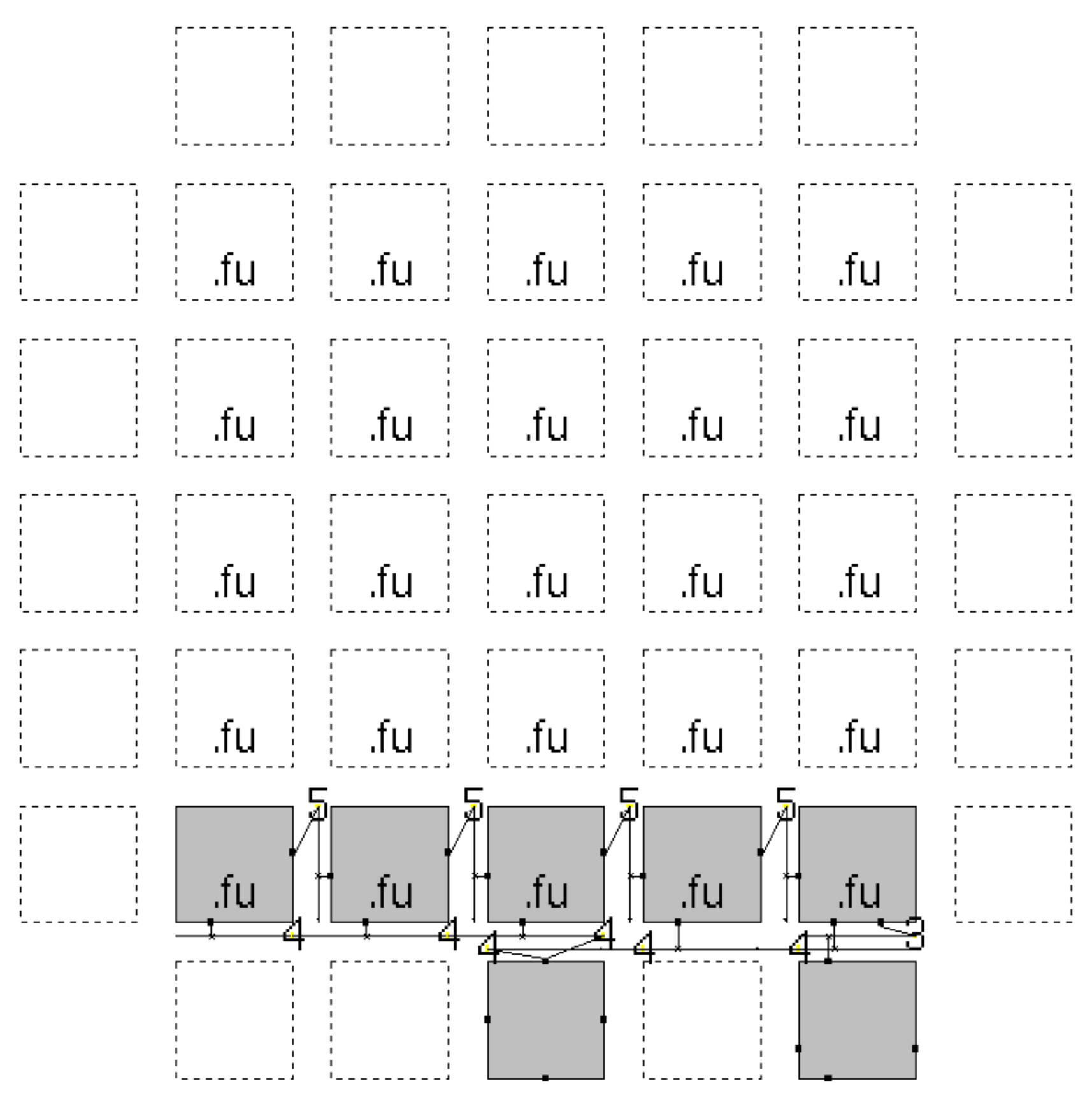}
		\label{par_1d}
	}
	\hfill
	\subfigure[FU-aware DFG where FU consists of two DSP blocks]{
		\centering
		\begin{tikzpicture}[>=latex',line join=bevel, scale = 0.28]
		\tiny%
		\node (N9) at (46.0bp,10.5bp) [draw=black,ellipse] {O0\_N7};
		\node (N1) at (112.0bp,466.5bp) [draw=black,ellipse] {I0\_N1};
		\node (N2) at (46.0bp,67.5bp) [draw=black,ellipse] {mul\_N2};
		\node (N3) at (102.0bp,238.5bp) [draw=black,text width=1.3cm,ellipse] {mul\_N3,    mul\_add\_Imm\_5\_N6};
		%\node (N4) at (112.0bp,409.5bp) [draw=black,ellipse,fill=red!20] {N4};
		\node (N5) at (112.0bp,352.5bp) [draw=black,text width=1.5cm, ellipse] {mul\_Imm\_16\_N4, mul\_sub\_Imm\_20\_N5};
		%\node (N6) at (102.0bp,181.5bp) [draw=black,ellipse,fill=red!20] {N6};
		
		%\draw [->] (N1) ..controls (146.27bp,446.62bp) and (178.55bp,423.54bp)  .. (167.0bp,399.0bp) .. controls (160.57bp,385.33bp) and (147.79bp,374.36bp)  .. (N5);
		%\draw [->] (N1) ..controls (159.02bp,444.04bp) and (224.0bp,407.12bp)  .. (224.0bp,353.5bp) .. controls (224.0bp,353.5bp) and (224.0bp,353.5bp)  .. (224.0bp,294.5bp) .. controls (224.0bp,243.78bp) and (166.06bp,209.78bp)  .. (N6);
		\draw [->] (N3) ..controls (102.0bp,211.14bp) and (102.0bp,211.14bp)  .. (N2);
		%\draw [->] (N4) ..controls (112.0bp,391.91bp) and (112.0bp,382.14bp)  .. (N5);
		\draw [->] (N5) ..controls (112.0bp,325.14bp) and (112.0bp,325.14bp)  .. (N3);
		\draw [->] (N1) ..controls (64.984bp,444.04bp) and (0.0bp,407.12bp)  .. (0.0bp,353.5bp) .. controls (0.0bp,353.5bp) and (0.0bp,353.5bp)  .. (0.0bp,180.5bp) .. controls (0.0bp,144.98bp) and (20.045bp,107.55bp)  .. (N2);
		\draw [->] (N1) ..controls (84.093bp,450.73bp) and (65.284bp,437.6bp)  .. (57.0bp,420.0bp) .. controls (31.453bp,365.71bp) and (34.192bp,340.5bp)  .. (N3);
		\draw [->] (N2) ..controls (46.0bp,49.908bp) and (46.0bp,40.144bp)  .. (N9);
		\draw [->] (N1) ..controls (112.0bp,448.91bp) and (112.0bp,439.14bp)  .. (N5);
		%\draw [->] (N6) ..controls (98.983bp,163.91bp) and (97.208bp,154.14bp)  .. (N2);
		%
		
		\end{tikzpicture}
		\label{fu_dfg_2d}
	}
	\hfill
	\subfigure[FU-aware DFG placed and routed on 5x5 overlay]{
		\centering
		\includegraphics[width=3.5cm]{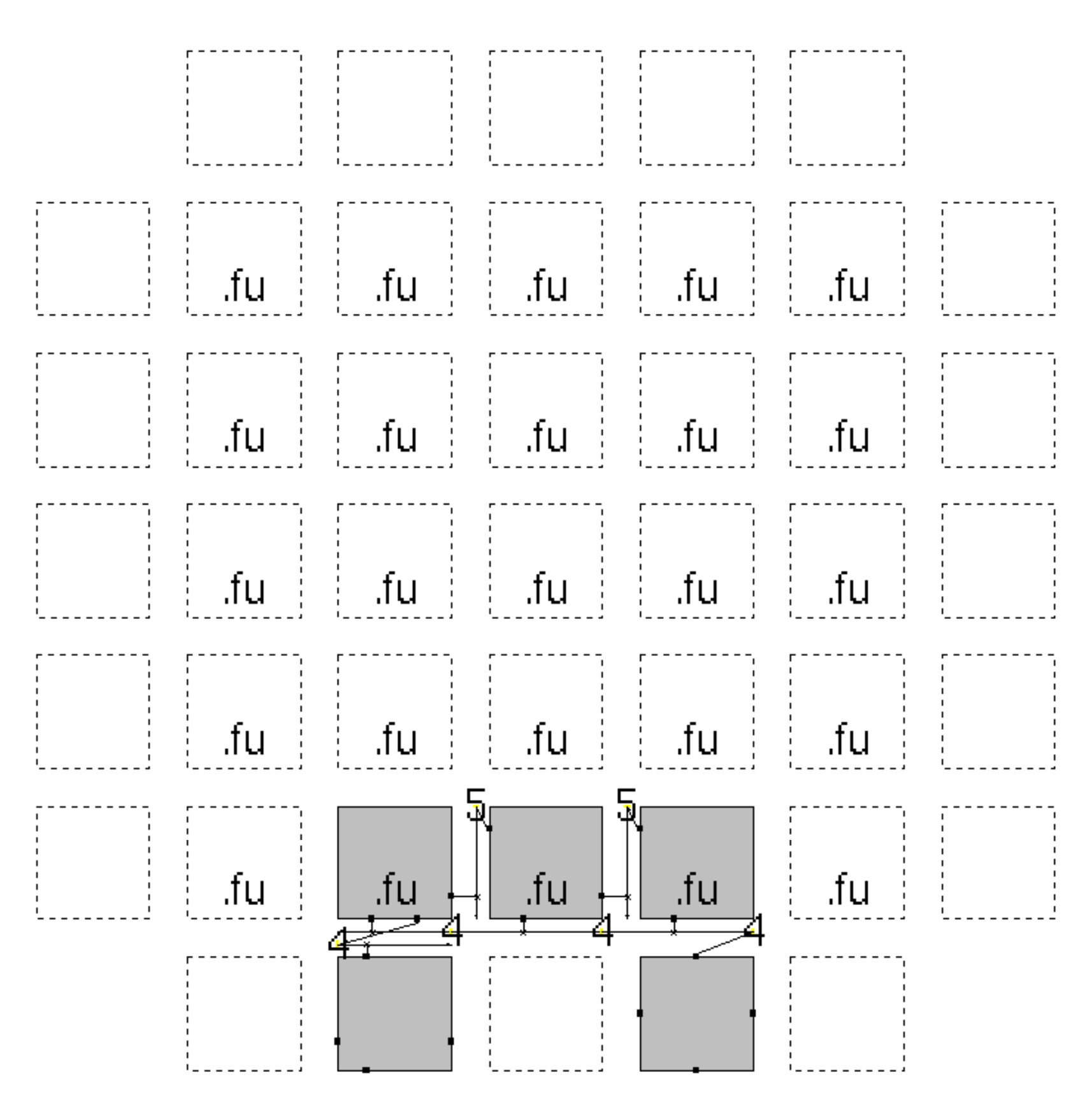}
		\label{par_2d}
	}
	
	\caption[]{FU aware mapping, placement and routing on overlay.}
	\label{dfgs}
\end{figure*}
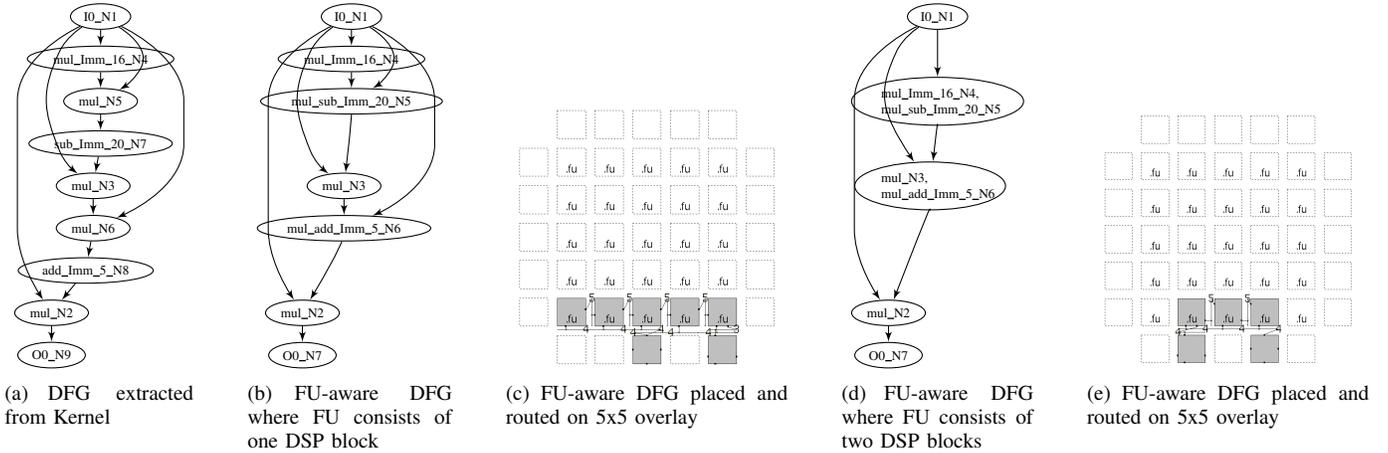

\subsubsection{DFG extraction from the optimized IR}
Starting with an optimized IR description of the compute kernel, our IR parser transforms this to a DFG description, given in Table~\ref{dfg.tbl}(a).
The DFG consists of nodes that represent operations and edges that represent the flow of data between operations.
A node executes in a pipelined manner, performs its operation and produces an output only when all of its inputs are ready, as per the dataflow model of computation.
Fig.~\ref{dfg} shows the representation of the example DFG described by Table~\ref{dfg.tbl}(a).

\subsection{DFG to FU-aware DFG transformation}
In this step, the DFG description is parsed and translated into an FU-aware DFG, as shown in Fig.~\ref{fu_dfg_1d}.
This involves merging nodes that can be combined into a single FU, based on the capabilities of the DSP block primitive.
For example, we can use the DSP block's multiply-subtract and multiply-add capabilities to collapse the N5--N7 and N6--N8 nodes in Fig.~\ref{dfg} into the N5 and N6 nodes of Fig.~\ref{fu_dfg_1d}, respectively.
As a result, the FU aware mapping requires only 5 FUs instead of the 7 that would be required if each node were mapped to a single FU, as occurs in many other overlays.
This results in the FU-aware DFG shown in Fig.~\ref{fu_dfg_1d}.

Furthermore, FUs can be made more complex by incorporating multiple DSP blocks to better balance resource usage~\cite{date2016-jain}.
For example using two DSP blocks within an FU, N4 and N5 can be combined together and similarly N3 and N6 can be combined together, resulting in another FU-aware graph as shown in Fig.~\ref{fu_dfg_2d}.

\subsection{Resource-aware FU netlist generation}
The FU-aware DFG for the kernel is replicated the appropriate number of times to fit the available resources as exposed by the OpenCL runtime.  
This replicated DFG is used to generate the FU netlist in VPR netlist format.
The replication factor used in this example is one, which means just one copy of the kernel would be mapped onto the overlay.

\subsection{Placement and routing of the FU Netlist}
VPR is then used to map DFG nodes onto the homogeneous FUs and DFG edges to the overlay routing paths to connect the mapped FUs.
Thus, rather than map logic functions to LUTs and single-bit wires to 1-bit channels, we map nodes in the graph to FUs, and 16-bit wires to 16-bit channels.
Fig.~\ref{par_1d} shows the DFG of Fig.~\ref{fu_dfg_1d} mapped on a 5$\times$5 overlay using the VPR place and route tool.
Similarly, Fig.~\ref{par_2d} shows the DFG of Fig.~\ref{fu_dfg_2d} mapped on a 5$\times$5 overlay (with two DSP blocks per FU).

\subsection{Latency Balancing}
Correct functioning of the mapped compute kernel is ensured only if it is latency balanced, which means that all FU inputs arrive at the same execution cycle.
As a result, the FUs have delay chains which must be configured to match these input latencies.
To determine the latency imbalance at each node, we parse the PAR output files and generate an overlay resource graph which is used to generate the configuration data (including the delay configuration) for the overlay. This information is loaded onto the overlay at runtime using the OpenCL API.

%-------------------------------------------------------------------------
% Section 4
%-------------------------------------------------------------------------
\section{Experiments}
\label{sec6_experiments}
\begin{figure}[!h]
	\centering
	\includegraphics[width=6.8cm]{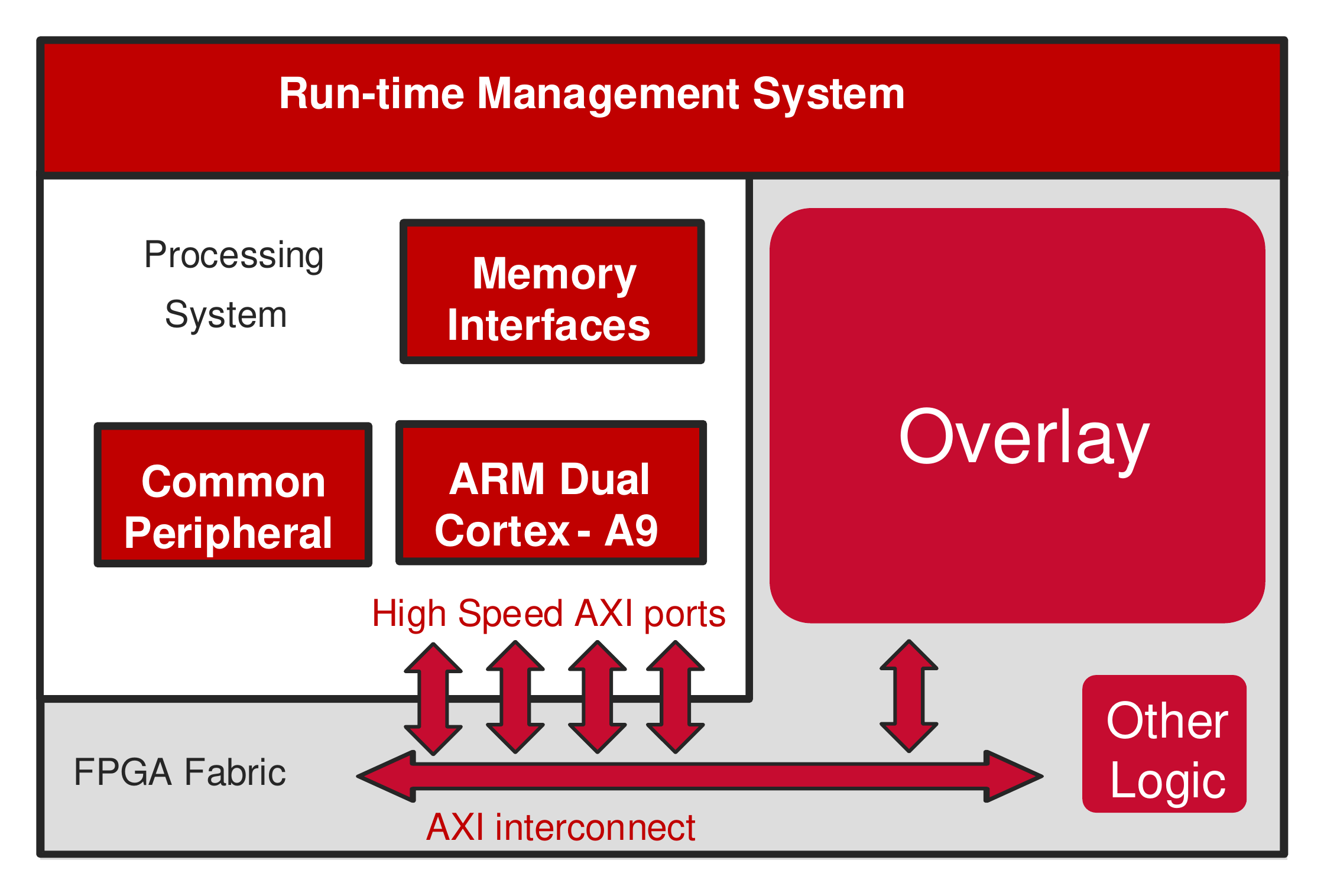}
	\caption{Overlay infrastructure, implemented on the Zynq, consisting of an Overlay whose size and FU type can be exposed by OpenCL runtime.} 
	\label{system}
\end{figure}

\begin{figure*}[!t]
	\centering
	\includegraphics[width=15cm]{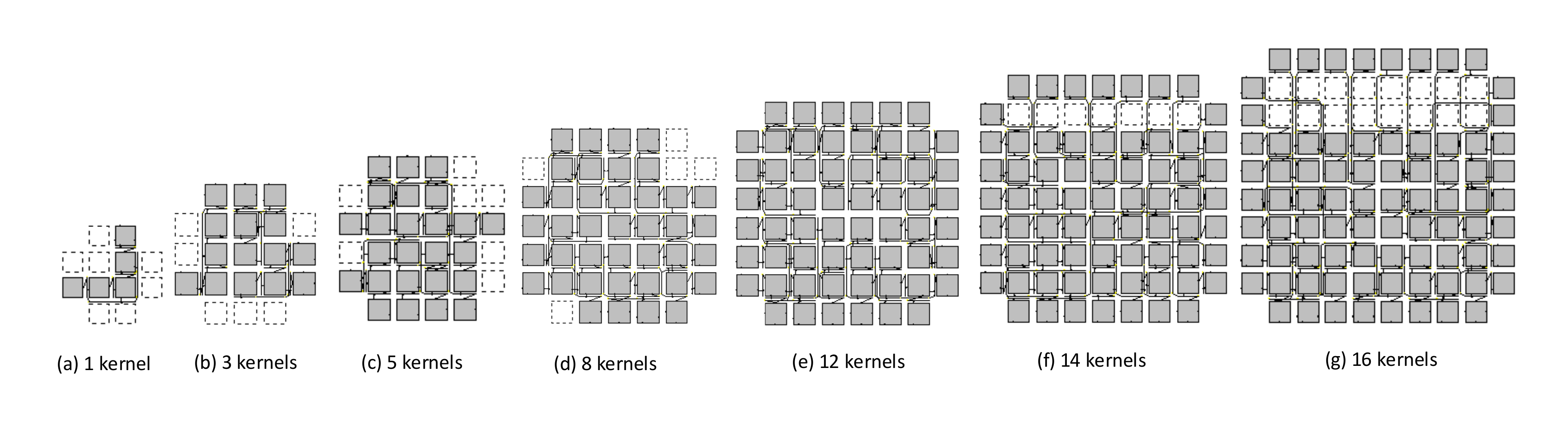}
	\caption{Performance scaling by the compiler using overlay size information provided by the OpenCL runtime.} 
	\label{scaling}
\end{figure*}

To demonstrate our on-demand OpenCL source-level compilation infrastructure, we consider a lightweight heterogeneous computing system based on the Xilinx Zynq XC7Z020, which consists of a dual-core ARM Cortex-A9 CPU, running at 667 MHz with 512 MB of RAM, and an Artix-7 FPGA fabric. 
The heterogeneous infrastructure, shown in Fig.~\ref{system}, includes the overlay in the programmable logic region of the Zynq FPGA.
The overlay size and FU type are exposed by the OpenCL runtime to the compiler so that it can dynamically replicate a suitable number of kernel copies to fully utilize the available overlay resources.
We deliberately do not consider a fixed overlay size as there may be situations in which other logic in the system consumes significant resources.
In that case, the overlay size can be changed and the fabric reconfigured incorporating this additional logic and the new appropriately sized overlay, without requiring any change to the OpenCL source code. 
For example, in the case where the other logic is large, leaving only minimal resources for a small, say 2$\times$2, overlay, this information can be exposed by the OpenCL runtime to the compiler which can then choose to map only a single copy of a kernel, as shown in Fig.~\ref{scaling}(a).
In the case where the other logic is minimal and most of the fabric resources can be used to map the overlay, we can fit an 8$\times$8 overlay and the compiler can then choose to map multiple copies of the kernel, in this case 16 copies of the \emph{Chebyshev} benchmark as shown in Fig.~\ref{scaling}(g), limited only by the available I/O.
Figs.~\ref{scaling}(b)--\ref{scaling}(f) show the cases in between.

\pgfdeclarelayer{background}
\pgfdeclarelayer{foreground}
\pgfsetlayers{background,main,foreground}

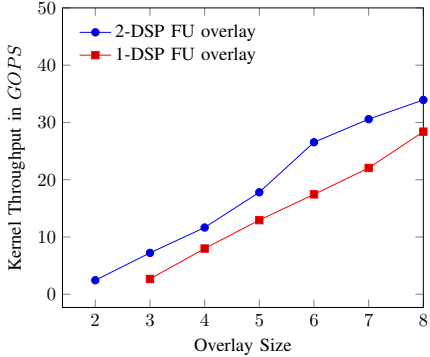
\begin{figure}[!h]
	\centering
	\subfigure{
		\centering
		\begin{tikzpicture}[scale=0.7]
		\begin{axis}[
		xlabel= Overlay Size,
		ylabel = Kernel Throughput in $GOPS$,
		ymax = 50,
		xmax = 8,
		legend pos=north west,
		legend style={draw=none}
		]
		\addplot plot coordinates {
			(2,     1*350*7/1000)
			(3,     3*344*7/1000)
			(4,     5*333*7/1000)
			(5,     8*318*7/1000)
			(6,     12*316*7/1000)
			(7,     14*312*7/1000)
			(8,     16*303*7/1000)
		};
		\addplot plot coordinates {
			%			(2,     2.6)
			(3,     1*382*7/1000)
			(4,     3*380*7/1000)
			(5,     5*370*7/1000)
			(6,     7*356*7/1000)
			(7,     9*350*7/1000)
			(8,     12*338*7/1000)
		};		
		
		\legend{2-DSP FU overlay\\1-DSP FU overlay\\}
		\end{axis}
		\end{tikzpicture}
		\label{resources_2}
	}
	
	\caption[]{Performance scaling by performing \emph{Chebyshev} kernel replication on different overlays}
	\label{scaling_perf}
	%\end{minipage}
	
\end{figure}

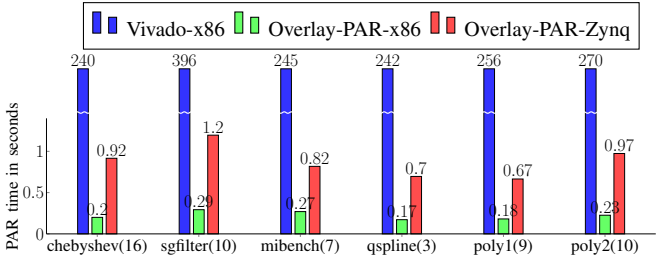
\begin{figure}[!t]
	\centering
	\begin{tikzpicture}[scale = 0.27]
	\begin{axis}[
	every axis plot post/.style={/pgf/number format/fixed},
	every axis label/.append style={font=\Huge},
	ybar=5pt,
	bar width=15pt,
	x=5cm,
	ymin=0,
	axis on top,
	ymax=1.4,
	xtick=data,
	ylabel={PAR time in seconds},
	enlarge x limits=0.1,
	symbolic x coords={chebyshev(16),sgfilter(10),mibench(7),qspline(3),poly1(9),poly2(10)},
	x tick label style={font=\Huge},
	y tick label style={font=\Huge},
	restrict y to domain*=0:2.0, % Cut values off at 14
	visualization depends on=rawy\as\rawy, % Save the unclipped values
	after end axis/.code={ % Draw line indicating break
		\draw [ultra thick, white, decoration={snake, amplitude=1pt}, decorate] (rel axis cs:0,1.05) -- (rel axis cs:1,1.05);
	},
	nodes near coords={%
		\pgfmathprintnumber{\rawy}% Print unclipped values
	},
	every node near coord/.append style={font=\Huge},
	%nodes near coords align={vertical},
	axis lines*=left,
	clip=false
	]
	\addplot[draw=black,fill=blue!80, draw opacity=1] coordinates {(chebyshev(16),240) (sgfilter(10),396) (mibench(7),245) (qspline(3),242) (poly1(9),256) (poly2(10),270)}; \label{workstation-vivado}
	\addplot[draw=black,fill=green!60, draw opacity=1] coordinates {(chebyshev(16),0.200) (sgfilter(10),0.293) (mibench(7),0.270) (qspline(3),0.172) (poly1(9),0.181) (poly2(10),0.226)}; \label{workstation-overlay}
	\addplot[draw=black,fill=red!70, draw opacity=1] coordinates {(chebyshev(16),0.916) (sgfilter(10),1.196) (mibench(7),0.818) (qspline(3),0.696) (poly1(9),0.665) (poly2(10),0.974)}; \label{zynq-overlay}
	\end{axis}
	\node [draw,fill=white] at (rel axis cs: 0.6,1.8) {\shortstack[l]{
			\ref{workstation-vivado} {\footnotesize Vivado-x86} \ref{workstation-overlay} {\footnotesize Overlay-PAR-x86} \ref{zynq-overlay} {\footnotesize Overlay-PAR-Zynq}}};

	\end{tikzpicture}
	\caption{Comparison of PAR times (in seconds).}
	\label{compile-time}
\end{figure}

Clearly, different sized overlays have different performance characteristics.
Fig.~\ref{scaling_perf} shows the effect of performance scaling using kernel replication on overlays having different sizes. The top (blue) curve shows the throughput in GOPS for replicated copies of the \emph{Chebyshev} kernel mapped to an overlay with two DSP blocks per FU. 
A throughput of $\approx35$ GOPS, representing 30\% of the peak overlay throughput, can be achieved using an $8\times8$ overlay by replicating 16 kernel instances while a single instance of the kernel provides a performance of 2.45 GOPS (representing 30\% of the peak performance of a $2\times2$ overlay).
The bottom (red) curve shows the throughput in GOPS for replicated copies of the \emph{Chebyshev} kernel mapped to an overlay with a single DSP block per FU. 
A throughput of $\approx28$ GOPS, representing 43\% of the peak overlay throughput of 65 GOPS, can be achieved using an $8\times8$ overlay by replicating 12 kernel instances while a single instance of the kernel achieves a performance of 2.66 GOPS (25\% of the peak performance of the $3\times3$ overlay).
The main benefit of resource aware replication of kernels at runtime is the possibility of exploiting higher performance on a larger FPGA fabric without changing the application source code.

\begin{table*}[!t]
	\renewcommand{\arraystretch}{1.3}
	\caption{Comparison of overlay implementations and direct FPGA implementations.}
	\label{results_final}
	\centering
	%\tiny
	\scriptsize
	%\footnotesize 
	\begin{tabular}{c|ccc|ccc|c|c|c}
		\toprule
		Benchmark		&\multicolumn{3}{c|}{Overlay implementations}&\multicolumn{3}{c|}{Direct FPGA implementations}& 					&				& \\
		\cline{2-7}			                                                                           	
		name			& PAR time		& $F_{max}$		& Resource		& PAR time		& $F_{max}$		& Resource		& Resource Penalty			& $F_{max}$			& PAR	\\
		& (seconds)		& (MHz)		& (DSP | Slices)& (seconds)		& (MHz)		& (DSP | Slices)& (DSP | Slices)			& Improvement	& speedup		\\		
		\midrule	  	                                                                                        	                                	
		chebyshev(16) 	& 0.2			& 300			& 128 | 12617				& 240			& 225		& 48 | 251		& 2.6$\times$ | 50$\times$  & 1.3$\times$   & 1200$\times$ \\
		sgfilter(10) 	& 0.29			& 300			& 128 | 12617				& 396			& 185		& 100| 797		& 1.2$\times$ | 15$\times$  & 1.6$\times$   & 1365$\times$ \\
		mibench(7)   	& 0.27			& 300 		    & 128 | 12617	& 245			& 230		& 21 | 403		& 6.0$\times$ | 31$\times$  & 1.3$\times$   & 907$\times$  \\
		qspline(3)   	& 0.17			& 300			& 128 | 12617				& 242			& 165		& 36 | 307		& 3.5$\times$ | 41$\times$  & 1.8$\times$   & 1423$\times$ \\
		poly1(9)     	& 0.18			& 300			& 128 | 12617				& 256			& 175		& 36 | 425		& 3.5$\times$ | 29$\times$  & 1.7$\times$   & 1422$\times$ \\
		poly2(10)     	& 0.23			& 300			& 128 | 12617				& 270			& 172		& 40 | 453		& 3.2$\times$ | 27$\times$  & 1.7$\times$   & 1173$\times$ \\
		
		\bottomrule	                                                                       
		
	\end{tabular}
\end{table*}

The time taken to map these different accelerator kernels to the FPGA is of major importance, and the traditional FPGA flow is too time consuming to enable the on-demand kernel replication, exploited with great success by GPUs, feasible.
To demonstrate the difference in the PAR times for our overlay we compile a set of benchmarks described in OpenCL onto the overlay (with two DSP blocks per FU) and measure the PAR time. 
This is then compared to that of a traditional FPGA implementation.
In this experiment we consider three different scenarios, with the results shown in Fig.~\ref{compile-time}.
Here, the number of replicated copies of the benchmark is shown in brackets after the benchmark name.
The first (blue) bar shows the PAR time using Vivado 2014.2 running on a HP Z420 workstation with an Intel Xeon E5-1650 v2 CPU running at 3.5 GHz with 16 GB of RAM, targeting the Zynq FPGA fabric. This is referred to as \emph{Vivado-x86}.
The second (green) bar, referred to as \emph{Overlay-PAR-x86}, shows the PAR time when the proposed overlay-based approach is used. Here our customized PAR tool flow is running on the same HP Z420 workstation as previously, and represents the situation where an FPGA accelerator card is installed into a workstation.
The third (red) bar, referred to as \emph{Overlay-PAR-Zynq}, shows the PAR time when the PAR tool is running on the Zedboard consisting of a Zynq XC7Z020 device having a dual-core ARM Cortex-A9 CPU, running at 667 MHz with 512 MB of RAM. 
Xillinux-1.3 is used as an operating system running on the dual-core ARM with Portable Computing Language (pocl) infrastructure~\cite{jaaskelainen2015pocl} installed.

For the set of 6 benchmarks, PAR takes on average 275\,s, 0.22\,s, and 0.88\,s, for \emph{Vivado-x86}, \emph{Overlay-PAR-x86}, and \emph{Overlay-PAR-Zynq}, respectively. This represents a speed-up in the workstation-based PAR process targeting the overlay of approximately 1250$\times$, compared to that of Vivado. 
When using the Zynq ARM processor, the PAR process targeting the overlay is still in excess of 300$\times$ faster.

While it is fairly obvious that a coarse-grained overlay will perform better than a fine-grained FPGA in terms of accelerator mapping speed, overlays do come with a significant resource and performance overhead. 
In an attempt to better quantify these overheads in relation to the use of OpenCL for on-demand source level compilation to the FPGA based overlay, we examine the performance metrics for the same six replicated benchmarks from Fig.~\ref{compile-time}. These are mapped to an 8$\times$8 overlay with two DSPs per FU using our tool chain, and directly to the FPGA fabric using Vivado 2014.2.
We examine the PAR time, the accelerator maximum frequency ($F_{max}$) and the FPGA resource utilisation (in terms of DSP blocks and logic slices) for the two implementations, as shown in Table~\ref{results_final}.
The benefits, or otherwise, of the overlay approach are shown at the right of the table. 
Here we see that for the 6 benchmarks, there is a resource penalty with an average increase of 3.4$\times$ in the DSP usage and 32$\times$ in the logic slice utilisation.
The overlay shows an improvement in the maximum frequency of 1.6$\times$ and in the PAR time of 1250$\times$.
Additionally, there is a significant difference in the configuration data sizes for the two implementations, and hence the configuration time.
The 8$\times$8 overlay requires 1061 bytes of data to fully configure it, taking 42.4\,$\mu$s, while the FPGA has a configuration data size of 4 Mbytes, taking 31.6\,ms to configure. This represents an improvement in the overlay configuration time of approximately 750$\times$.

\section{Conclusion and Future Work}
\label{sec7_conclusion}
We have presented an approach for runtime compilation of OpenCL kernels onto coarse-grained overlays for improving accelerator design productivity.
The methodology benefits from the high level of abstraction afforded by using the OpenCL programming model, while the mapping to overlays offers fast compilation in the order of seconds, even on an embedded processor.
We demonstrate an end-to-end compilation flow with resource aware mapping of kernels to the overlay.
Using a typical workstation, the overlay place and route is $\approx$1250 times faster than the FPGA place and route using Vivado 2014.2. Furthermore, the overlay can be reconfigured in less than 50$\mu$s using the OpenCL API.
We successfully installed the pocl infrastructure on the Zedboard to support execution of OpenCL applications, and demonstrated place and route onto the overlay running entirely on the Zynq.
However, the area overheads associated with the overlay are significantly higher than for a direct FPGA implementation, with an increase of 3.4$\times$ and 32$\times$ in DSP and logic slice usage, respectively.
We are currently examining techniques to reduce the overlay resource penalty so that the full benefits of on-demand OpenCL kernel source compilation can be realised.

\def\IEEEbibitemsep{1pt plus 1pt}
\bibliographystyle{IEEEtran}
\bibliography{IEEEabrv,references}

\begin{thebibliography}{10}
\providecommand{\url}[1]{#1}
\csname url@samestyle\endcsname
\providecommand{\newblock}{\relax}
\providecommand{\bibinfo}[2]{#2}
\providecommand{\BIBentrySTDinterwordspacing}{\spaceskip=0pt\relax}
\providecommand{\BIBentryALTinterwordstretchfactor}{4}
\providecommand{\BIBentryALTinterwordspacing}{\spaceskip=\fontdimen2\font plus
\BIBentryALTinterwordstretchfactor\fontdimen3\font minus
  \fontdimen4\font\relax}
\providecommand{\BIBforeignlanguage}[2]{{%
\expandafter\ifx\csname l@#1\endcsname\relax
\typeout{** WARNING: IEEEtran.bst: No hyphenation pattern has been}%
\typeout{** loaded for the language `#1'. Using the pattern for}%
\typeout{** the default language instead.}%
\else
\language=\csname l@#1\endcsname
\fi
#2}}
\providecommand{\BIBdecl}{\relax}
\BIBdecl

\bibitem{chen2012invited}
D.~Chen and D.~Singh, ``Invited paper: Using {OpenCL} to evaluate the
  efficiency of {CPUs}, {GPUs} and {FPGAs} for information filtering,'' in
  \emph{Proceedings of the International Conference on Field Programmable Logic
  and Applications ({FPL)}}.\hskip 1em plus 0.5em minus 0.4em\relax IEEE, 2012,
  pp. 5--12.

\bibitem{gao2014characterization}
S.~Gao and J.~Chritz, ``Characterization of {OpenCL} on a scalable {FPGA}
  architecture,'' in \emph{Proceedings of the International Conference on
  ReConFigurable Computing and FPGAs (ReConFig)}, 2014, pp. 1--6.

\bibitem{stone2010opencl}
J.~E. Stone, D.~Gohara, and G.~Shi, ``{OpenCL:} a parallel programming standard
  for heterogeneous computing systems,'' \emph{Computing in science \&
  engineering}, vol.~12, no. 1-3, pp. 66--73, 2010.

\bibitem{singh2013harnessing}
D.~P. Singh, T.~S. Czajkowski, and A.~Ling, ``Harnessing the power of {FPGAs}
  using altera's {OpenCL} compiler,'' in \emph{Proceedings of the International
  Symposium on Field Programmable Gate Arrays ({FPGA)}}.\hskip 1em plus 0.5em
  minus 0.4em\relax ACM, 2013, pp. 5--6.

\bibitem{ahmad201616}
S.~Ahmad, V.~Boppana, I.~Ganusov, V.~Kathail, V.~Rajagopalan, and R.~Wittig,
  ``A 16-nm multiprocessing system-on-chip field-programmable gate array
  platform,'' \emph{IEEE Micro}, vol.~36, no.~2, 2016.

\bibitem{owaida_synthesis_2011}
M.~Owaida, N.~Bellas, K.~Daloukas, and C.~Antonopoulos, ``Synthesis of
  {Platform} {Architectures} from {OpenCL} {Programs},'' in \emph{{IEEE}
  Symposium on Field-Programmable Custom Computing Machines ({FCCM)}}, May
  2011.

\bibitem{czajkowski_opencl_2012}
T.~Czajkowski, U.~Aydonat, D.~Denisenko, J.~Freeman, M.~Kinsner, D.~Neto,
  J.~Wong, P.~Yiannacouras, and D.~Singh, ``From {OpenCL} to high-performance
  hardware on {FPGAs},'' in \emph{Proceedings of the International Conference
  on Field Programmable Logic and Applications (FPL)}, Aug. 2012, pp. 531--534.

\bibitem{shagrithaya_enabling_2013}
K.~Shagrithaya, K.~Kepa, and P.~Athanas, ``Enabling development of {OpenCL}
  applications on {FPGA} platforms,'' in \emph{Proceedings of the International
  Conference on Application-Specific Systems, Architecture Processors
  ({ASAP)}}, Jun. 2013, pp. 26--30.

\bibitem{stitt_intermediate_2011}
G.~Stitt and J.~Coole, ``Intermediate fabrics: Virtual architectures for
  near-instant {FPGA} compilation,'' \emph{IEEE ESL}, vol. 3(3), pp. 81--84,
  2011.

\bibitem{capalija_high-performance_2013}
D.~Capalija and T.~S. Abdelrahman, ``A high-performance overlay architecture
  for pipelined execution of data flow graphs,'' in \emph{Proceedings of the
  International Conference on Field Programmable Logic and Applications
  ({FPL)}}, 2013.

\bibitem{benson2012design}
J.~Benson, R.~Cofell, C.~Frericks, C.-H. Ho, V.~Govindaraju, T.~Nowatzki, and
  K.~Sankaralingam, ``Design, integration and implementation of the {DySER}
  hardware accelerator into {OpenSPARC},'' in \emph{International Symposium on
  High Performance Computer Architecture ({HPCA)}}, 2012.

\bibitem{jainadapting2016}
A.~K. Jain, X.~Li, S.~A. Fahmy, and D.~L. Maskell, ``Adapting the {DySER}
  architecture with {DSP} blocks as an overlay for the {Xilinx} {Zynq},''
  \emph{SIGARCH Computer Architecture News}, vol.~43, no.~4, pp. 28--33, 2016.

\bibitem{fccm2015-jain}
A.~K. Jain, S.~A. Fahmy, and D.~L. Maskell, ``Efficient {Overlay} architecture
  based on {DSP} blocks,'' in \emph{{IEEE} Symposium on {FPGAs} for Custom
  Computing Machines ({FCCM)}}, 2015.

\bibitem{date2016-jain}
A.~K. Jain, D.~L. Maskell, and S.~A. Fahmy, ``Throughput oriented {FPGA}
  overlays using {DSP} blocks,'' in \emph{Proceedings of the Design, Automation
  and Test in Europe Conference ({DATE)}}, 2016.

\bibitem{jain2016deco}
A.~K. Jain, X.~Li, P.~Singhai, D.~L. Maskell, and S.~A. Fahmy, ``{DeCO:} a
  {DSP} block based {FPGA} accelerator overlay with low overhead
  interconnect,'' in \emph{{IEEE} Symposium on Field-Programmable Custom
  Computing Machines ({FCCM)}}, 2016, pp. 1--8.

\bibitem{ebeling_rapid_1996}
C.~Ebeling, D.~C. Cronquist, and P.~Franklin, ``{RaPiD} - reconfigurable
  pipelined datapath,'' in \emph{Field-Programmable Logic Smart Applications,
  New Paradigms and Compilers}, 1996, pp. 126--135.

\bibitem{liang_smartcell:_2009}
C.~Liang and X.~Huang, ``{SmartCell:} an energy efficient coarse-grained
  reconfigurable architecture for stream-based applications,'' \emph{EURASIP
  Journal on Embedded Systems}, vol. 2009, no.~1, pp. 518--659, Jun. 2009.

\bibitem{mei_adres:_2003}
B.~Mei, S.~Vernalde, D.~Verkest, H.~D. Man, and R.~Lauwereins, ``{ADRES:} an
  architecture with tightly coupled {VLIW} processor and coarse-grained
  reconfigurable matrix,'' in \emph{Field Programmable Logic and Application},
  Jan. 2003, pp. 61--70.

\bibitem{jain2016coarse}
A.~K. Jain, D.~L. Maskell, and S.~A. Fahmy, ``Are coarse-grained overlays ready
  for general purpose application acceleration on {FPGAs}?'' in
  \emph{Proceedings of the International Conference on Pervasive Intelligence
  and Computing}.\hskip 1em plus 0.5em minus 0.4em\relax IEEE, 2016.

\bibitem{rashid2014comparing}
R.~Rashid, J.~G. Steffan, and V.~Betz, ``Comparing performance, productivity
  and scalability of the {TILT} overlay processor to {OpenCL HLS},'' in
  \emph{Proceedings of the International Conference on Field-Programmable
  Technology ({FPT)}}, 2014.

\bibitem{coole2014fast}
J.~Coole and G.~Stitt, ``Fast, flexible high-level synthesis from {OpenCL}
  using reconfiguration contexts,'' \emph{IEEE Micro}, vol. 34(1), 2014.

\bibitem{cloudcom2015-fahmy}
S.~A. Fahmy, K.~Vipin, and S.~Shreejith, ``Virtualized {FPGA} accelerators for
  efficient cloud computing,'' in \emph{Proceedings of the International
  Conference on Cloud Computing Technology and Science (CloudCom)}, 2015.

\bibitem{jsps2014-jain}
A.~K. Jain, K.~D. Pham, J.~Cui, S.~A. Fahmy, and D.~L. Maskell, ``Virtualized
  execution and management of hardware tasks on a hybrid {ARM-FPGA} platform,''
  \emph{J. Signal Process. Syst.}, vol. 77(1--2), 2014.

\bibitem{liu2015quickdough}
C.~Liu, H.-C. Ng, and H.~K.-H. So, ``{QuickDough:} a rapid {FPGA} loop
  accelerator design framework using soft {CGRA} overlay,'' in
  \emph{Proceedings of the International Conference on Field-Programmable
  Technology ({FPT)}}, 2015, pp. 56--63.

\bibitem{jaaskelainen2015pocl}
P.~J{\"a}{\"a}skel{\"a}inen, C.~S. de~La~Lama, E.~Schnetter, K.~Raiskila,
  J.~Takala, and H.~Berg, ``pocl: A performance-portable {OpenCL}
  implementation,'' \emph{International Journal of Parallel Programming},
  vol.~43, no.~5, pp. 752--785, 2015.

\end{thebibliography}

\end{document}